**The actual pressure and temperature at the melt of elemental vanadium**

Joseph Gal* and Lonia Friedlander

Ilse Katz Institute for Nanoscale Science and Technology,
Ben-Gurion University of the Negev, Beer Sheva 84105, Israel

Keywords: High pressure, Melting curve, Equation of state, Shock Waves, Vanadium

PACS numbers: 61.50.Ks, 62.50.+p, 64.30+t, 71.15Nc

*jgal@bgu.ac.il

# Abstract

It is claimed that all of the pressure scales of the reported melting curves derived by diamond anvil cell experiments require a correction which takes into account the pressure thermal shift, where vanadium is an illustrative example. The linear behavior of the thermal pressure ($P_{th}$) vs. the temperature, as predicted by first principles theoretical assumptions is then experimentally confirmed. This allows extrapolation to determine of the actual pressure and thermal temperature at the melt. Accounting for the role of the pressure transmitting media in diamond anvil cell experiments, the analysis of elemental vanadium melting curve is presented. It is shown that the appropriate correction of shock waves melting data which takes into account the radiation absorbed by the LiF window, applies only to vanadium metal. The correct pressure scale of vanadium metal as derived by diamond anvil cell is presented.

1. **Introduction**

During the last five decades, diamond anvil cells (DAC) have been frequently used to determine the equations of state (EOS) and melting curves of elemental metals [1]. However, DAC experimental melting curve data have never matched, neither the shock waves (SW) experimental results nor theoretical density functional theory (DFT) simulations [2,3,4]. In a recent publication, Y. Zhang et al. [5] took into account the radiation absorption by the LiF window; thus correcting the reported SW data of elemental vanadium. This approach is strongly supported by the first principles DFT-Z method, proving that this procedure is indeed reasonable for elemental Vanadium. In addition, however, Y. Zhang et al. suggested that this procedure should also be applied to all the d-electron transition metals (see [5], conclusions). Zhang's suggestion neglects the fact that melting curves, derived by DAC, are strongly dependent on the pressure transmitting medium (PTM) [4,6,7].

In a previous paper, it was argued that isochoric conditions exist in the DAC's chamber [4]. Namely, increasing the temperature in the DAC's chamber provokes an additional thermal pressure. In practice, a mechanical pressure gauge to measure this thermal pressure does not exist. Therefore, the actual pressure must be estimated, either from the pressure shift of each melting point relative to the initial ambient pressure, or from first principles calculations of the lattice component in the P-V-T equation of state [8]. However, the direct determination of the thermal pressure from experimental results was not available for vanadium until the results reported by Y. Zhang. In the present contribution, the linear behavior of $P_{th}$ vs. the temperature, predicted by first principles assumptions, is confirmed. This allows the extrapolation of the thermal pressure and the thermal temperature up to melt for elemental V.

Many experiments have shown that the melting curves of the same element measured by laser-heated DAC (LH-DAC) in different experiments, reveal different melting points using different PTMs [6,7]. The pressure medium is expected to distribute the pressure homogeneously within the pressure chamber, preventing non-hydrostatic effects such as, pressure gradients, shear stress, or inhomogeneous pressure. Upon increasing the temperature at each pressure, however, the examined sample and the PTM both are subject to increases of their volumes. However, the volume expansion is suppressed by the chamber's finite volume, provoking an increase in the thermal pressure over the whole system. In some cases, the PTM melts and remains liquid during the whole experiment. Therefore, the pressure reported does not necessarily represent the actual pressure experienced by the sample. The effects of the PTM on the measured sample, and the PTM's response to P,T changes throughout the experiment must be taken into account. Thus, the pressure scale of the reported melting curves and the equations of state (EOS) isotherms should be corrected according to the actual pressure in the cell, and V is an illustrative example.

**2. Melting curve of elemental V**

The corrected melting curves of elemental V, derived by SW and DAC experiments, are depicted in Fig. 1(a). The non-modified uncorrected SW data and the DAC measurements are compared with the corrected DAC and SW data as reported by Y. Zhang et al. [5]. The black asterisks and the magenta solid line represent the corrected SW data, taking into account the radiation absorption of the LiF window. The SW data without this correction are also shown for comparison. The role of the different PTMs in the DAC measurements, taken from Errandonea et al. [7], are clearly displayed in the figure. The red squares are the first DAC measurements of V utilizing an Ar PTM [9]. The blue squares are DAC measurements by Errandonea et al. [7] of metallic V, using a NaCl PTM [7,10]. The green diamonds are V embedded between two dried KCl layers [5,11,12]. The fits to the melting points were performed by the third order Birch-Murnaghan (BM) EOS under the constraint of the combined approach [4,13] (Appendix 2. below), where the Grüneisen parameter $\gamma_o$ is a fitting parameter, and marked in the figure as $B_o/B_o'/\gamma_o$.

The SW data at the pressure region 148-225 GPa (red circles) were reported by E. Errandonea et al. [7]. Both, the blue and magenta solid lines present Gilvarry-Lindemann approximations [13, see Appendix]. The black asterisks are the corrected SW data, according to Y. Zhang et al [5] and in accordance agreement with the DFT-Z method (green solid line).

The experimental melting points of V embedded in a KCl PTM (green diamonds) [5], are compared to V embedded in a NaCl [10] PTM are also depicted in Fig. 1(b). The melting curves of KCl and NaCl are shown in the figure (red and black solid lines, respectively) clearly demonstrate that V is embedded in solid KCl, while the

V is embedded in a liquid NaCl PTM. The magenta solid line presents the anchor (combined approach) for the corrected SW measurements (see Fig. 1(a)) of V.

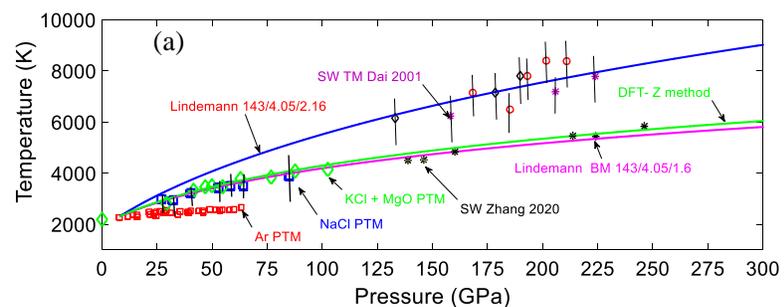

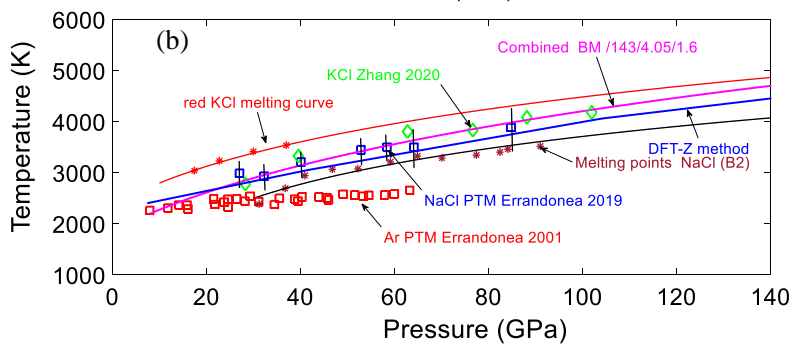

**Fig. 1**: (a): The corrected melting curve of vanadium metal (magenta solid line) is confirmed by first principles using the DFT-Z method (green solid line). The non-modified SW data and the DAC measurements are compared to the corrected DAC and SW data reported by Y. Zhang et al. [5], (black asterisks). Comparing the melting points derived by LH-DAC using an Ar PTM (red squares), compared to those derived using NaCl and KCl PTMs (blue squares and green diamonds, respectively), clearly demonstrates the role of the PTM in DAC experiments. (b): Melting curves of KCl [11,12] and NaCl [10] (red and black lines) shown with respect to the experimental melting points of elemental vanadium. The derived melting points of V, embedded in liquid NaCl (blue squares) PTM are compared to V embedded in solid KCl (green diamonds) PTM. The magenta solid line presents the (combined approach) anchor for the corrected SW measurements depicted above in (a).

### 3. Vanadium: Actual pressure in LH-DAC

The melting curve of vanadium reported by Y. Zhang et al. was determined by in-situ x-ray diffraction in a LH-DAC and is depicted in Fig. 2 [5, therein Fig.4]. The magenta and blue solid lines present the melting curve of V fitted with the Gilvarry-Lindemann criterion, which is confirmed by the first principles simulations (DFT-Z method). The colored asterisks represent the thermal pressure shifts as a function of the applied temperature at each pressure [5]. The dashed black lines represent the 300K starting applied pressures. The angle (difference) between the dashed lines and the asterisk points, indicates that indeed thermal pressure ($P_{th}$) is provoked by increasing the temperature, thus confirming the assumption that isochoric conditions exist in DAC chambers containing solid PTM. The horizontal green double arrow lines indicate the pressure thermal shifts ($P_{th}$) pinched to the melt, where the temperature increase is marked by short green solid line.

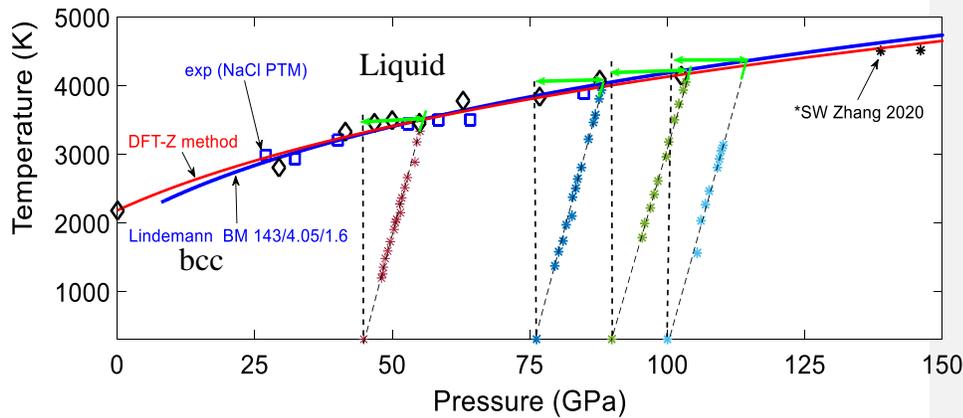

**Fig. 2:** Thermal pressure at several applied pressures of vanadium metal. The red and blue solid lines represent the melting curve of V fitted with the combined approach approximation (see appendix), which is further confirmed by DFT-Z method simulation. The blue squares are the experimental melting points derived LH-DAC with V embedded in a liquid NaCl PTM. The black diamonds represent V embedded in a solid KCl PTM. The 100 GPa melting point is V embedded in a MgO PTM [7]. The colored asterisks are the thermal pressure shifts as a function of the applied temperature, reported by Y. Zhang [5]. The dashed black lines present the 300K initial applied pressures. The angle between the dashed lines and the asterisk points, indicates the thermal pressure ($P_{th}$) provoked by the increase of the temperature; while the horizontal green double arrow lines indicate the ($P_{th}$) thermal shifts ($P_{th}$). The bright green solid lines demonstrate the further increase of the temperature while crossing the melting curve.

## 4. Thermal pressure shift vs. the temperature

Based on Y. Zhang et al. [5] experimental data of Fig. 2, the linear behavior of the thermal pressure upon increasing the applied temperature can't be missed. The experimental pressure shifts upon elevating the temperature under applied pressure are shown distinctly in Fig. 3. The red thick solid line represents the theoretical lattice contribution derived by the approximated P-V-T equation of state [8] revealing;

$P(V,T) = P_o + P_{th} = P_o + 3.2 \cdot 10^{-3} (T-300)$ GPa (see Appendix).

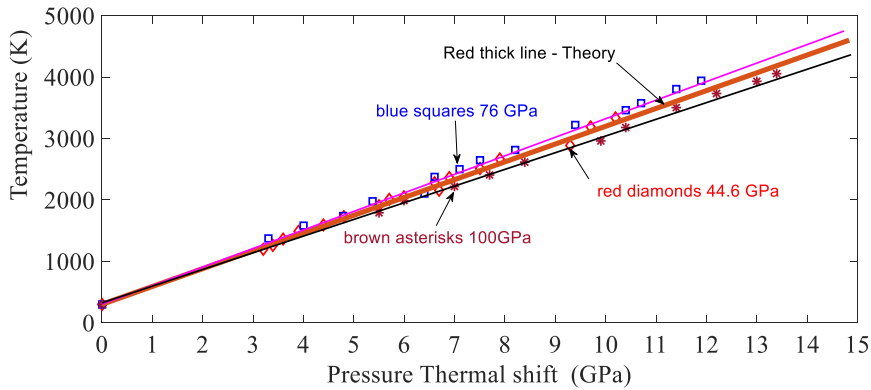

Fig. 3: Vanadium; based on Y. Zhang et al. [5] experimental data, the pressure shift upon elevating the temperature under applied pressure of 44.6 GPa (red diamonds) 76 GPa (blue squares) and 100 GPa (brown asterisks). The red thick line represents the theoretical lattice contribution derived from the P-V-T equation of state [8] (see Appendix).

5. **Discussion**

The contribution of Y. Zhang at al. [5] clearly demonstrates that all the experimental SW and the DAC data reported to-date, need corrections. As shown, there is no discrepancy between the static DAC and the dynamic SW measurements of V, if the calculated thermal correction to the cold experimental data and the radiation absorption by the LiF window, are taken correctly into account. The proposed approach is indeed confirmed by first principles, using the *ab-initio* DFT-Z method. simulation.

Unfortunately, the conclusion of Y. Zhang et al. quote " This study provides us a sound methodology of combined static and dynamic experiments and theoretical approaches to determine the phase diagrams and melting curves of d-orbital transition metals and their compounds at ultrahigh P-T " is absolutely miss leading, as each d-orbital transition metal shows different behavior of the DAC experimental melting data relative to SW data and relative to the DFT and QMD first principles simulations [4,2,14]. Nevertheless, the experimental results reported by Y. Zhang et al. are of high quality and should be awarded.

Y. Zhang et al. did not mention or accounted for the differences between the liquid NaCl and the solid KCl PTMs used in all of the many reported V experiments (see Fig. 1(b)). Therefore, this procedure applies only to LH-DAC experiments utilizing solid PTM, or to low compressible PTMs, such as $Al_2O_3$ [15] and MgO [16]. The same phenomenon have been reported by A. Karandiker and R. Boehler on Ta [17] , as the flushing laser-speckle pattern on the sample surface forms a heat spike, in which the molten Ta expansion is suppressed by surrounding solid Ta metal provoking an additional thermal pressure yielding higher pressures and elevated melting temperatures.

The actual pressure in the LH-DAC chamber can be estimated from first principles calculations utilizing the P-V-T equation of state [8] (see Appendix 2. below). However, the use of experimental data to directly determine the actual pressure close to the melt in a LH-DAC experiment, has not, up to now, been possible. In the present contribution, a method to directly derive the thermal pressure and the thermal temperature pinched to the melting curve is proposed, using vanadium as an instructive example. The corrected scale (actual pressure) of V melting curve is depicted in Fig.4.

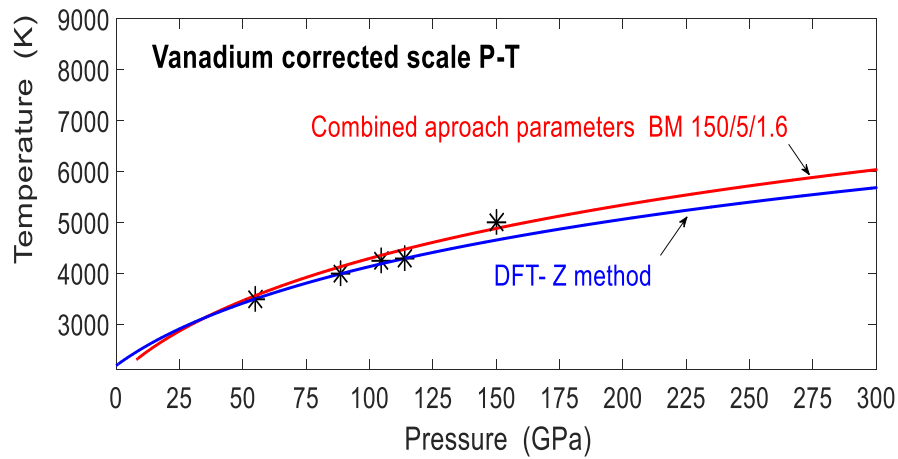

**Fig.4**: Corrected scale melting curve of elemental vanadium derived by extrapolation (see Fig.2), confirmed by DFT-Z method.

The analysis of any experimental point is referred to the center of mass of the error bar. What should be looked at is the trend of the experimental points relative to the fitting line. Thus, the corrected scale melting curve of V shown in Fig.4 is different than that claimed by Y. Zhang et al. [5].

Moreover, Ir [18] and Pt [19] metals confirm the linear behavior of the thermal shift as a function of the applied temperature, where the pressure shift ($P_{th}$) at the melt can be derived by extrapolation.

## 6. Conclusions

1. All of the pressure scales reporting melting curves derived by DAC experiments need correction, taking into account the measured thermal shifts.

2. The linear behavior of the thermal pressure ($P_{th}$) vs. the temperature, as predicted by first principles theoretical assumptions, is experimentally confirmed allowing extrapolation to determine the actual pressure and temperature at the melt. Vanadium serves as an illustrative example.

3. The correction of the SW experimental data accounting for the radiation absorbed by the LiF window (SW data) is applicable only to V. The claim that the correction of the radiation absorption by the LiF window should be applied to all d-orbitals transition metals is miss-leading.

4. The advantage of the Gilvarry-Lindeman constrained fitting procedure (combined approach) of the measured melting points is demonstrated; in this procedure the volume, the bulk moduli and $\gamma_o$ are simultaneously derived.

**Acknowledgement**

The authors gratefully acknowledge Prof. Z. Zinamon, Department of Particle Physics, Weizmann Institute of Science, Rehovot – Israel, for the many helpful and illuminating discussions and comments.

# Appendix

1. **Actual pressure in the P-V-T space**

First-principles (*ab-initio*) [8] lead to the following relation:

$$P(V,T) = P_c + \gamma_{lattice} \, C_{v\,lattice} \, V^{-1} \, [T-T_o+E_o/C_{v\,lattice}] + \tfrac{1}{4} \, r_o \, \gamma_e \, \beta_o \, (\rho/\rho_o)^{1/2} \, T^2 \quad (1)$$

Here $P_c$ is the cold pressure, $C_v$ is the lattice specific heat above $T_o$. $T_o$ is the ambient temperature. $C_{v,lattice}$ is taken as a constant (usually at room temperature, following the approximation of Altshuler et al. [8,21]), $E_o$ is the lattice thermal internal energy at $T_o$ ($E_o = \int_0^T Cv \, dT$)), and $\gamma_{lattice}$ is the lattice Grüneisen parameter. V is the volume. For the case of extreme temperatures, $\gamma_e$ is the electronic Grüneisen parameter and $\beta_o$ is the electronic specific heat coefficient. At low temperatures, equation (1) is usually expressed as $P(V,T) = P_c + P_{th} + P_e$. Assuming that $C_v \gamma/v$ is a constant [8] and that there is a negligible contribution of from $E_o/C_v$, one arrives with a simplified equation for the lattice contribution: $P_{th} = C \, (T-300)$ where C is a fitting parameter (see Fig.3).

2. **Gilvarry-Lindemann approximation: The Combined Approach.**

According to Lindenmann's criterion, the melting temperature $T_m$ is related to the Debye temperature $\Theta_D$ as follows; (assuming a Debye solid):, $T_m = C\, V^{2/3}\, \Theta_D^2$, where V is the volume and C is a constant to be derived for each specific metal. Assuming that $\gamma = \gamma_o\, (r_o/r)^q$ and $q = 1$, one gets the approximation :

$$T_m(V) = T_{mo}\, (V/V_o)^{2/3}\, \exp[2\,\gamma_o\, (1 - V/V_o)] \qquad (2)$$

Combining this with the third-order Birch–Murnaghan (BM) equation of state:

$$P(V) = 3/2\, B_o\, [\, (V_o/V)^{7/3} - (V_o/V)^{5/3}\, ]\, [1 + 3/4\, (B_o' - 4)\, \{\, (V_o/V)^{2/3} - 1\, \}] \qquad (3)$$

We apply the constrain that the fitting parameter V of the experimental equation of state (EOS) at ambient temperature, must simultaneously fit (eqs. 2 and 3) the experimental melting and the EOS data where $\gamma_o$ is a fitting parameter.

For the reader who wants to use the combined approach the relevant Matlab program for the vanadium case follows:

```
dB=B'
B=Bo
V=13.9:-0.086:4
Vo=13.9;
VVo=V/13.9;
A=1.5*B*(VVo.^-2.333)-(VVo.^-1.666);
C=1-(0.75*(4-dB));
D=C*((VVo.^-0.6666)-1);
E=D.*C;;
P=E.*A;
r=VVo;
Gama=1.63
EX=exp(2*Gama*(1-r));
Tm=2300*r.^0.6666;
TM=EX.*Tm;
plot(P,TM,'r')
plot(P,V,'b')
```

Here B=$B_o$, dB=$B_o'$, Gama= $\gamma_o$, is a fitting parameter, TM is the melting curve according to the Gilvarry-Lindemann criterion.

### 3. Lindemann-Gilvarry vs. Simon-Glatzel fitting procedure of melting curves:

The Simon-Glatzel equation is in fact a combination of the Murnaghan EOS and Lindemann's criterion: $Tm = T_{ref} (B_o' P - P_{ref}/B_o + 1)^{2(\gamma - 1/3 + f)}$ where f is a coefficient in Mores potential. In the case of vanadium; $Tm \sim 2183 (1+P/32)^{0.46}$ [7]. The three parameters 2183, 32 and 0.46 do not contribute in the derivation of γ and f constants.

The advantage of the constrained Gilvarry-Lindeman procedure is that the volume, the bulk moduli and $\gamma_o$ are simultaneously derived.